\documentclass[10pt,draft]{dis03}
\usepackage{epsf,amsmath}

\begin{document}

\title{HEAVY QUARKONIUM PRODUCTION WITH POLARIZED HADRONS AND PHOTONS%
\thanks{Work supported by DFG through Grants No.\ KL~1266/1-3 and KN~365/1-1,
by BMBF through Grant No.\ 05~HT1GUA/4, and by Sun Microsystems through
Academic Equipment Grant No.~EDUD-7832-000332-GER.}}

\author{MICHAEL KLASEN and BERND A.~KNIEHL \\
II. Institut f\"ur Theoretische Physik, Universit\"at Hamburg \\
Luruper Chaussee 149, 22761 Hamburg, Germany \\
E-mail: klasen@mail.desy.de}

\maketitle

\begin{abstract}
\noindent
We investigate the inclusive production of prompt $J/\psi$ mesons in polarized
hadron-hadron and photon-hadron collisions in the factorization formalism of
NRQCD. Numerical results are presented for BNL RHIC-Spin and the approved SLAC
fixed-target experiment E161 to assess the feasibility to access the
spin-dependent parton distributions in the polarized proton and photon. We
point out that data on $J/\psi$ production taken by the PHENIX
Collaboration in unpolarized proton-proton collisions at RHIC tend to favor
the NRQCD factorization hypothesis, while they significantly overshoot the
theoretical prediction of the CSM.
\end{abstract}

\section{Introduction}

Before the advent of the non-relativistic QCD (NRQCD) factorization formalism,
heavy quarkonium production with polarized proton or photon beams was believed
to provide reliable information on the spin-dependent gluon distributions
of the polarized proton or photon. At present, however, we are faced with the
potential problem that NRQCD predictions at lowest order have a
considerable normalization uncertainty due to the introduction of
non-perturbative color-octet matrix elements, which were not present in the
color-singlet model (CSM) and which also have to be extracted from experiment
\cite{Braaten:1996pv}.

In order to clarify the question if heavy quarkonium production with polarized
proton or photon beams remains to be a useful probe of the polarized gluon
densities, we investigate inclusive $J/\psi$ production in polarized $pp$ and
$\gamma p$ collisions at RHIC-Spin and E161. Other processes with the
potential to constrain the polarized gluon distributions include the production
of low-mass lepton pairs \cite{Berger:1999es}. Numerical results relevant for
$J/\psi$ production in polarized $\gamma\gamma$ collisions at DESY TESLA as
well as a complete list of analytical results for all polarized partonic cross
sections can be found in Ref.\ \cite{Klasen:2003zn}.

\section{\label{sec:num}\boldmath$J/\psi$ production at BNL RHIC-Spin and SLAC
E161}

\begin{figure}
\centerline{\epsfxsize=10cm\epsfbox{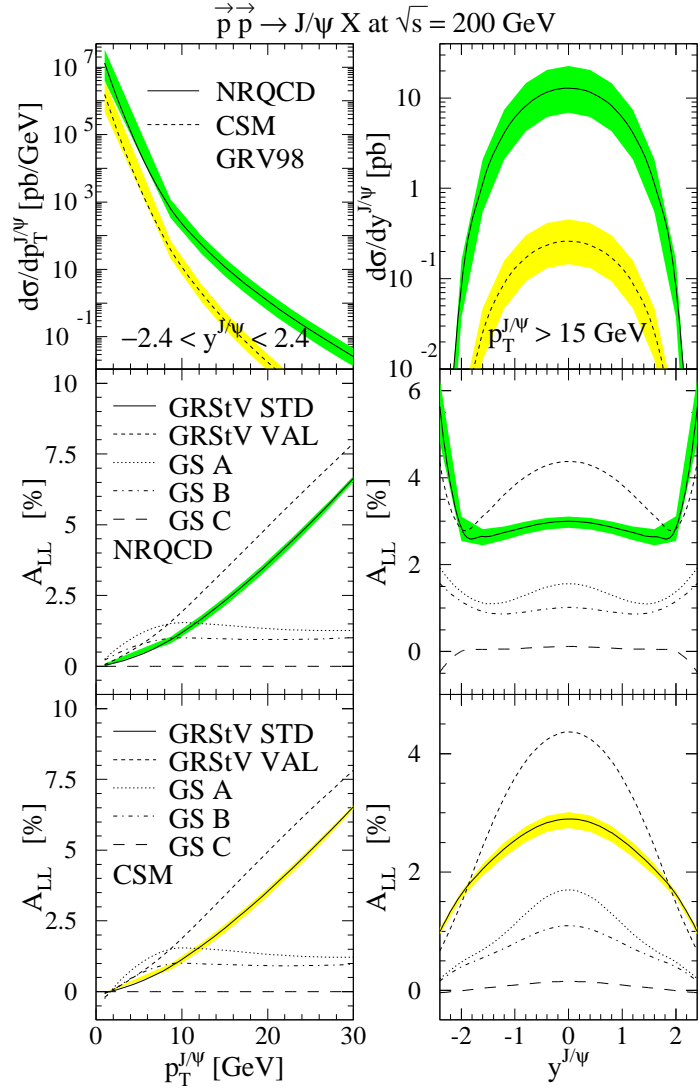}}
\caption[*]{Cross sections $d\sigma/dp_T$ and $d\sigma/dy$ (first panel) and
asymmetries ${\cal A}_{LL}$ (second and third panels) of $pp\to J/\psi+X$ at
RHIC-Spin with $\sqrt S=200$~GeV as functions of $p_T$ (left) and $y$ (right)
in NRQCD and the CSM. \label{fig:rhic2}}
\end{figure}

\begin{figure}
\centerline{\epsfxsize=10cm\epsfbox{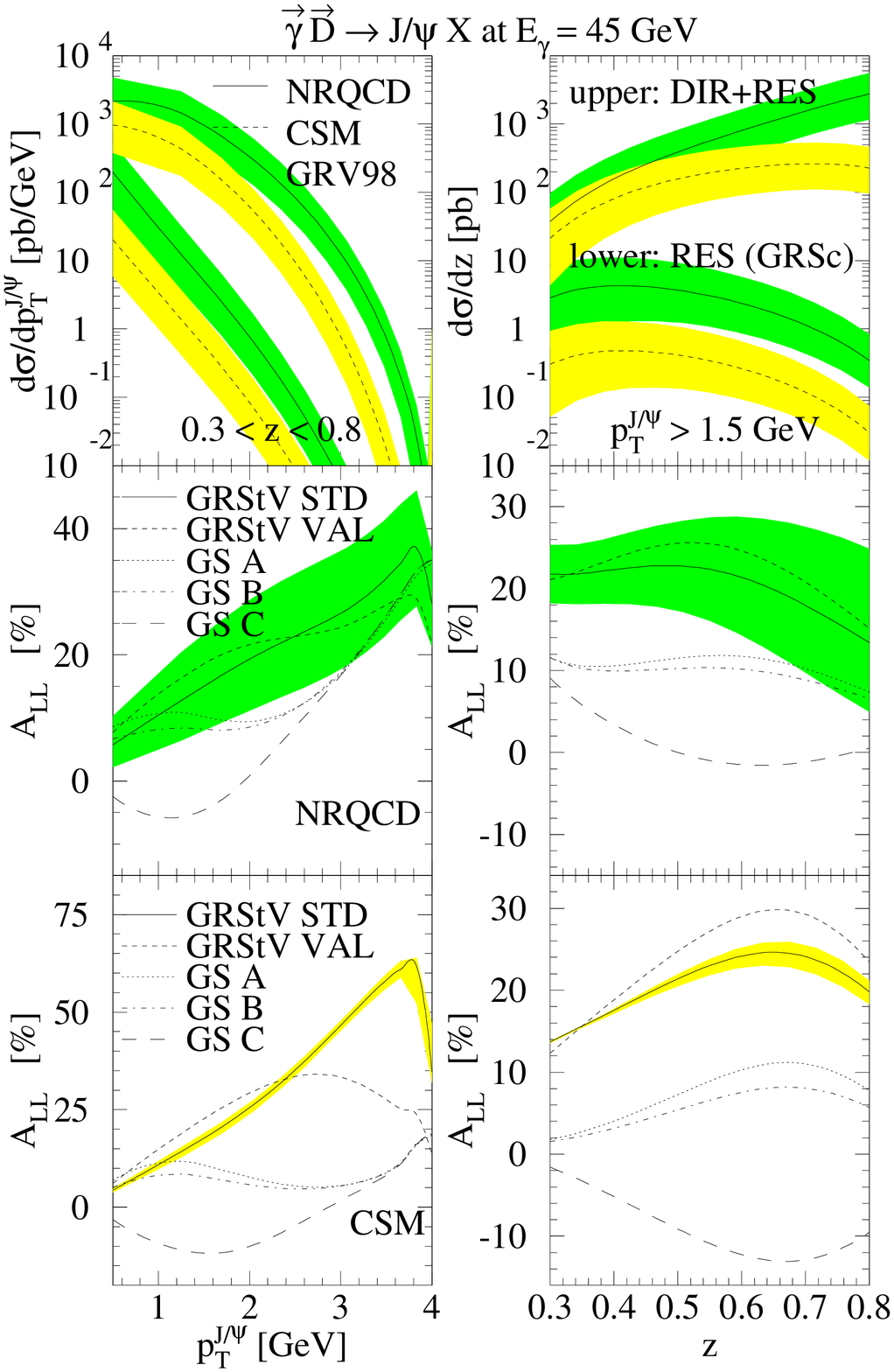}}
\caption[*]{Same as in Fig.~\ref{fig:rhic2}, but for $\gamma D\to J/\psi+X$ at
E161 with $E_\gamma=45$~GeV. Inelasticity $z$ is used instead of rapidity
$y$. In the first panel, the resolved-photon contributions are also shown
separately.
\label{fig:slac}}
\end{figure}

At RHIC, proton beams with longitudinal polarization of approximately 70\%
collide with center-of-mass energy up to $\sqrt{S}=500$ GeV and luminosity
${\cal L}=2\times10^{32}$~cm$^{-2}$s$^{-1}$. In E161, circularly polarized
photons with energies between 35 and 48 GeV will collide on a fixed target
made of longitudinally polarized deuterium \cite{Bunce:2000uv}.
In Figs.~\ref{fig:rhic2} and \ref{fig:slac}, NRQCD and CSM predictions for the
unpolarized cross sections $d\sigma/dp_T$ and $d\sigma/dy$ or $d\sigma/dz$
are displayed in the first panel, while those for the double longitudinal-spin
asymmetry ${\cal A}_{LL}$ are shown in the second and third panels. The $p_T$
distributions are integrated over the intervals $|y|<2.4$ and $0.3<z<0.8$,
respectively, while the $y$ or $z$ distributions are integrated over all
kinematically allowed values of $p_T$ in excess of 15 and 1.5 GeV. The shaded
bands indicate the theoretical uncertainties in the NRQCD and CSM default
predictions. As our default polarized parton densities, we employ the GRStV-STD
set in the proton and the GRSi-MAX set in the photon \cite{Gluck:2000dy} (for a
detailed discussion of our input parameters see
\cite{Klasen:2003zn}). We assume the ideal case of 100\% beam polarization.
Realistic polarization is accounted for by scaling ${\cal A}_{LL}$ with
$[P(p)]^2$ and $P(\gamma)P(D)$, respectively.

At RHIC-Spin, the differences in ${\cal A}_{LL}$ for various parton densities
are large against the combined theoretical uncertainties from other sources, so
that sufficiently precise measurements will increase our knowledge on the
spin-dependent parton structure of the polarized proton. The NRQCD and CSM
$p_T$-dependences incidentally almost coincide, so that the polarized proton
densities can be explored in a model-independent fashion. On the other hand,
the NRQCD and CSM predictions for the $y$ distribution exhibit strikingly
different shapes in the forward and backward directions.

At E161, the situation is less favorable since the theoretical uncertainties
are larger, due to the low photon-nucleon center-of-mass energy $\sqrt S=m_N(2
E_\gamma+m_N)\approx9.2$~GeV. This is especially the case for the NRQCD
prediction shown in the second panel of Fig.~\ref{fig:slac}. Nevertheless, it
should be possible to discriminate between the GRStV and GS sets. In the CSM,
the resolving power of ${\cal A}_{LL}$ used to be much  better, as is evident
from the third panel of Fig.~\ref{fig:slac}. In this sense, the introduction
of NRQCD, which was necessary to overcome phenomenological and conceptual
problems of the CSM, led to some aggravation.

\begin{figure}
\centerline{\epsfxsize=6.5cm\epsfbox{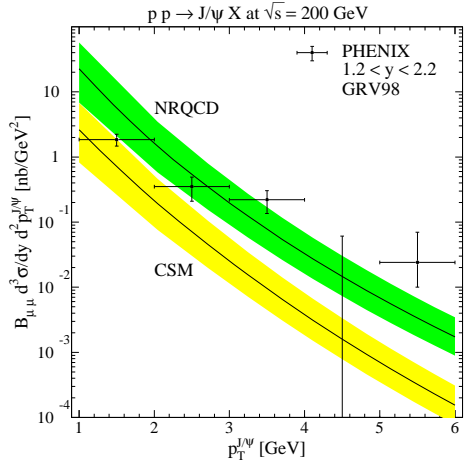}}
\caption[*]{PHENIX data for $B(J/\psi\to\mu^+\mu^-)d^3\sigma/dy\,d^2p_T$ at
RHIC with $\sqrt S=200$~GeV and in the interval $1.2<y<2.2$ as a function of
$p_T$ are compared with NRQCD and CSM predictions.
\label{fig:phenix}}
\end{figure}

As can be seen from the first panels in Figs.~\ref{fig:rhic2} and
\ref{fig:slac}, the normalization of the unpolarized cross section is a
distinctive discriminator between NRQCD and the CSM. In fact, data from the
PHENIX Collaboration \cite{Adler:2003qs} at RHIC, with $\sqrt S=200$~GeV, tend
to favor the NRQCD prediction compared to the CSM one (see
Fig.~\ref{fig:phenix}).
Since the $J/\psi$ mesons are tagged through their
decays to $\mu^+\mu^-$ pairs, the factor $B(J/\psi\to\mu^+\mu^-)=(5.88\pm0.10)
\%$ is included in the theoretical predictions.
We observe that, for $p_T>2$~GeV, the data are nicely described by the NRQCD
prediction, while they significantly overshoot the CSM one. In the bin
1~GeV${}<p_T<{}$2~GeV, the comparison has to be taken with a grain of salt
since the NRQCD prediction and the $P$-wave contribution to the CSM one suffer
from infrared and collinear singularities at $p_T=0$, which still feed into
that bin as an artificial enhancement.

\section{\label{sec:con}Summary}

For inclusive production of prompt $J/\psi$ mesons in polarized hadron-hadron
and photon-hadron collisions at RHIC and E161, we found that the spread in the
asymmetries for different parton densities in general considerably exceeds the
combined theoretical uncertainties from other sources, which we estimated
rather conservatively. Therefore, even within the NRQCD formalism these
experiments have discriminative power w.r.t.\ the spin structure of the
proton and photon. As a by-product, we found that PHENIX data of unpolarized
hadroproduction of $J/\psi$ mesons at RHIC favor NRQCD as compared to the CSM.
This is in line with previous findings in $pp$, $\gamma p$, and $\gamma\gamma$
collisions \cite{Braaten:1996pv,Klasen:2001mi}.

\section*{Acknowledgments}

We thank L.\ Mihaila and M.\ Steinhauser for their collaboration.


\begin{thebibliography}{0}

\bibitem{Braaten:1996pv}
E.~Braaten, S.~Fleming and T.~C.~Yuan,
Ann.\ Rev.\ Nucl.\ Part.\ Sci.\  {\bf 46} (1996) 197;
%
M.~Kr\"amer,
Prog.\ Part.\ Nucl.\ Phys.\  {\bf 47} (2001) 141;
%
M.~Klasen,
Rev.\ Mod.\ Phys.\  {\bf 74} (2002) 1221.

\bibitem{Berger:1999es}
E.~L.~Berger, L.~E.~Gordon and M.~Klasen,
Phys.\ Rev.\ D {\bf 62} (2000) 014014 and
%
RIKEN Rev.\  {\bf 28} (2000) 44.

\bibitem{Klasen:2003zn}
M.~Klasen, B.~A.~Kniehl, L.~N.~Mihaila and M.~Steinhauser,
hep-ph/0306080, Phys.\ Rev.\ D (to appear).

\bibitem{Bunce:2000uv}
G.~Bunce, N.~Saito, J.~Soffer and W.~Vogelsang,
Ann.\ Rev.\ Nucl.\ Part.\ Sci.\  {\bf 50} (2000) 525;
%
P.~Bosted,
SLAC-PUB-9269.

\bibitem{Gluck:2000dy}
M.~Gl\"uck, E.~Reya, M.~Stratmann and W.~Vogelsang,
Phys.\ Rev.\ D {\bf 63} (2001) 094005;
%
M.~Gl\"uck, E.~Reya and C.~Sieg,
Eur.\ Phys.\ J.\ C {\bf 20} (2001) 271.

\bibitem{Adler:2003qs}
S.~S.~Adler {\it et al.}  [PHENIX Collaboration],
hep-ex/0307019.

\bibitem{Klasen:2001mi}
M.~Klasen, B.~A.~Kniehl, L.~Mihaila and M.~Steinhauser,
Nucl.\ Phys.\ B {\bf 609} (2001) 518 and
%
Phys.\ Rev.\ Lett.\  {\bf 89} (2002) 032001.


\end{thebibliography}
\end{document}